\documentclass[pre,twocolumn,amsmath,showpacs,superscriptaddress,floatfix]{revtex4}
\usepackage{graphicx}

\begin{document}
\title{Universal and non-universal properties of transitions to
spatio-temporal chaos in coupled map lattices.}

\author{ Ren\'e Mikkelsen}

\affiliation{ Center for Chaos and Turbulence Studies, The Niels Bohr
Institute, Blegdamsvej 17, DK-2100, Copenhagen {\O}, Denmark}

\author{Martin van Hecke}

\affiliation{ Kamerlingh Onnes Laboratory, University of Leiden, Niels
Bohrweg 2, 2333 CA, Leiden, The Netherlands.  }

\author{Tomas Bohr}

\affiliation{Department of Physics, The Danish Technical University,
DK-2800 Kgs. Lyngby, Denmark }

\date{\today}

\begin{abstract}
We study the transition from laminar to chaotic behavior in
deterministic chaotic coupled map lattices and in an extension of
the stochastic Domany--Kinzel cellular automaton ~\cite{DK}. For
the deterministic coupled map lattices we find evidence that
``solitons'' can change the {\em nature} of the transition: for
short soliton lifetimes it is of second order, while for longer
but {\em finite} lifetimes, it is more reminiscent of a first
order transition. In the second order regime the deterministic
model behaves like Directed Percolation with infinitely many
absorbing states; we present evidence obtained from the study of
bulk properties and the spreading of chaotic seeds in a laminar
background. To study the influence of the solitons more
specifically, we introduce a soliton including variant of the
stochastic Domany--Kinzel cellular automaton. Similar to the
deterministic model, we find a transition from second to first
order behavior due to the solitons, both in a mean field analysis
and in a numerical study of the statistical properties of this
stochastic model. Our study illustrates that under the appropriate
mapping some deterministic chaotic systems behave like stochastic
models; but it is hard to know precisely which degrees of freedom
need to be included in such description.
\end{abstract}

\maketitle

\section{Introduction.}

Spatiotemporal chaos (STC) occurs in many spatially extended
deterministic systems and remains notoriously difficult to
characterize \cite{CATS}. Therefore one may attempt to map such
deterministic chaotic systems onto stochastic models for which many
more analytical methods are available.  It is then tacitly assumed
that, after sufficient coarse graining of the deterministic model, the
role of deterministic chaos can be taken over by the noise in the
stochastic system. A critical test of the validity of such mappings
are the predictions for the transitions between qualitatively
different states that extended chaotic systems display. The key
question is then: {\em are transitions in deterministic chaotic
systems governed by the universality classes of stochastic systems?}

As is known for a variety of spatiotemporal chaotic systems
\cite{CATS,local_struct} and as we will show below for the
deterministic system at hand, chaotic states in extended systems often
display a mixture of rather regularly propagating structures and more
disordered behavior. When the propagating structures, that we will
refer to as ``solitons'' (following \cite{Grass}) have a finite
lifetime, it may seem that they can be ignored after sufficient coarse
graining.  We will find strong indications that this is {\em not}
always the case, and we will give an example where their influence may
even be so strong as to change the nature of the transition.  We will
also show that extending simple stochastic models with the appropriate
solitonic degrees of freedom can mimic this behavior quite accurately:
not only can we change the order of the transition, we can also get
transient non-universal scaling of the type observed in coupled map
lattices \cite{CM}.  Therefore we conclude that, in many cases,
deterministic chaotic systems {\em can indeed} be mapped to stochastic
models. A short account of our work has already been published
\cite{prl}.

\subsection{Historical background}

Chat\'e and Manneville ~\cite{CM2} introduced the notion of a
universal transition to extended chaos via ``spatiotemporal
intermittency'' (STI) ~\cite{CM3,STI} in a study of the {\em
deterministic} damped Kuramoto-Sivashinsky partial differential
equation \cite{KS}. STI states are composed of ``turbulent'' (chaotic)
and ``laminar'' (ordered) patches, and the laminar patches remain so
except for contamination by turbulence at their boundaries. These
states are conjectured to occur quite generally when, locally, laminar
and turbulent dynamics are separated by a subcritical bifurcation, and
indeed a large number of different experimental systems and
theoretical models display STI \cite{STI}.

As a function of their parameters, STI systems display a transition
from states where the turbulence eventually dies out to states where
the turbulence spreads and dominates. Pomeau proposed ~\cite{Pomeau}
an analogy between this transition and the phase transition of the
stochastic process known as Directed Percolation (DP); for an
introduction to DP, see e.g.\ ~\cite{DP1,DP2}. In directed percolation
one considers the spreading of ``activity'' in an absorbing, inactive
background. Earlier, Grassberger ~\cite{Grassberger} and Janssen
~\cite{Janssen} had conjectured that any $stochastic$ process with an
unique absorbing state should be in the same directed percolation
universality class.

Relating laminar to inactive and turbulent to active states appears to
map spatiotemporal intermittency to directed percolation. To verify
whether {\it deterministic} chaotic models with an absorbing state
would be in the DP universality class, Chat\'e and Manneville
introduced a very simple coupled map lattice (CML) that displays STI
and numerically obtained the critical exponents that characterize the
transition from inactive to active states. Surprisingly, these
critical exponents appear to vary with the parameters and are in
general different from the DP values. Therefore the
Chat\'e--Manneville model appears to be {\it not} in the DP
universality class and not even universal.

Grassberger and Schreiber ~\cite{Grass} pointed out that the presence
of long lived traveling structures which they call ``solitons'' in the
Chat\'e--Manneville model may lead to large crossover times, and
conjectured that in the long-time limit the behavior of the
Chat\'e--Manneville model would be in the DP universality class.

Recently, the Chat\'e--Manneville model with an asynchronous update
rule was studied ~\cite{Tomas}. Here random sites are chosen to be
iterated forward while keeping the others unaltered. For this model,
the solitons observed for the standard synchronous update rule are
suppressed and the critical exponents are universal with DP values,
implying that the Chat\'e--Manneville model with asynchronous updating
belongs to the DP universality class. However, the asynchronous
updating introduces an element of stochasticity into the model, thus
ruining the deterministic character of the original model.

\subsection{Outline}

In this paper we will study a {\em deterministic} extension of the
Chat\'e--Manneville CML that facilitates the tuning of the soliton
properties. We will demonstrate that the influence of solitons may be
much more profound than setting a crossover time, since they appear to
be able to change the type of transition from second to first order.
The role of the solitons is further illustrated in an extension of the
{\em stochastic} Domany-Kinzel cellular automaton. In its standard
form, all sites of this model can be either active or inactive, but we
will add a ``solitonic'' degree of freedom that mimics the behavior of
the solitons in the CML's. The mean field equations of this stochastic
model show a transition from second order DP-like, behavior to a first
order transition when the soliton lifetimes are increased. Numerical
studies of this stochastic model also find evidence for such a
crossover to first order behavior, although it is very difficult to
asses the asymptotic behavior for our model. In any case, we present
strong numerical evidence that the transition is not an ordinary
second order transition and that there is no asymptotic scaling
regime, although there are appears to be a transient which displays
non-universal scaling behavior.

Our study illustrates that for extended systems it is a difficult task
to faithfully map a deterministic system to a stochastic counterpart.
In this particular case, localized propagating structures can be
identified as responsible for the breakdown of DP universality, but
one can imagine that less easily identified properties of the
deterministic dynamics could be responsible for such a breakdown in
other systems.

The outline of this paper is as follows. In section \ref{sCML} we
discuss the coupled map lattices. Starting from a brief discussion of
the classic Chat\'e-Manneville model, we introduce our extension to
lattices of two-dimensional maps in section \ref{ss2D}.  We show that
the new parameter that is introduced has a profound effect on the
importance of ``solitons'', and that long living solitons change the
transition from inactive to active states from a second to a first
order transition in section \ref{sssol}. In the second order regime,
we estimate the bulk critical exponents using finite size scaling
techniques in section \ref{ssfinite}, and measure spreading exponents
in section \ref{ssspread}. All this data is consistent with the
coupled map lattice being in the universality class of Directed
Percolation with infinitely many absorbing states, provided that
soliton lifetimes are short. In section \ref{sstoch} we discuss the
extension of the standard Domany-Kinzel cellular automata which
includes new degrees of freedom that mimic the solitons of the
coupled map lattices. The mean field equations for this model are
studied in section \ref{ssmft}, and these show a transition from
second to first order behavior as a function of the soliton lifetimes.
We study the phenomenology and its statistical bulk properties of the
full model in the soliton rich regime in section \ref{ssphe}. The
behavior of the model in the soliton rich regime is quite distinct
from an ordinary second order transition.

\section{Coupled Map Lattices.}\label{sCML}

The model introduced by Chat\'e and Manneville consists of coupled
maps, each of which either performs ``laminar'' or chaotic motion.
The model was motivated by the fact that studies of the deterministic
partial differential equations, such as the damped
Kuramoto-Sivashinsky equation, are numerically quite demanding and had
not provided enough precision to allow a definitive comparison to DP
\cite{CM,CM2}. In one spatial dimension their coupled map lattice was
defined according to
\begin{equation}\label{cm}
u_i(n+1)=f(u_i(n))+\frac{\varepsilon}{2}\Delta_f u_i(n)
\end{equation}
where the subscripts $i$ denote the spatial position, $n$ is the
discrete time and $\Delta_f u_i(n) = f(u_{i-1}(n)) - 2f(u_i(n)) +
f(u_{i+1}(n))$. This expression is a discrete approximation of
diffusive coupling in one dimension and introduces spatial
correlations in the system; the parameter $\varepsilon$ is a measure
of the coupling strength between a site $i$ and its two nearest
neighbors at sites $(i-1)$ and $(i+1)$.

The map $f$ is chosen such that locally the scalar field $u_i$ can be
in either of two states: the absorbing (laminar) or the chaotic
(turbulent) one. When $u<1$, $f$ is a standard tent map of the form
$f(u)=r(\frac{1}{2} - |u-\frac{1}{2}|)$ which displays chaotic
behavior, while in the region where $u>1$, $f$ is simply the identity
and leads to a laminar state. The sharp discontinuity in $f$ ensures
that the two states are distinguishable at each site.  The parameter
$r>2$ determines the steepness of the tent map as well as the
transition ratio from the chaotic to the laminar regime in the absence
of coupling.

The form of the diffusive coupling ensures that turbulent sites cannot
be spontaneously generated in a background of laminar sites: states
where all sites are laminar remain so, and the laminar state is truly
absorbing.
The laminar state is not unique: Updating a state where all sites
are in the laminar regime $(u_i >1)$ leads, via the diffusion
operator, to a state where all variables are equal to the global
average value $\bar{u}$.

Once initiated, turbulent activity can spread through this CML by
infecting laminar patches from their boundaries. The effectiveness of
the resulting spreading of the chaos depends on the values of $r$ and
$\varepsilon$.  Suppose we study the behavior of this system by
keeping $r$ fixed while varying the coupling strength $\varepsilon$.
Completely analogous to DP, a critical value
$\varepsilon=\varepsilon_c(r)$ exists, such that for
$\varepsilon<\varepsilon_c$ an absorbing state is reached with unit
probability while sustained chaotic behavior (in the thermodynamic
limit) is found for $\varepsilon>\varepsilon_c$.  Taking the density
of chaotic sites or ``activity'' $m$ as an order parameter,
transitions from a ``laminar'' state (where $m$ decays to zero) to a
``turbulent'' state (where $m$ reaches a finite value in an infinite
system) can be studied.

\subsection{Extensions to two-dimensional maps}\label{ss2D}

Coupled map lattice's can, in principle, be related to continuous time
physical systems of weakly coupled elements by interpreting the map
$f$ as a return map on a Poincar\'e section.  The time spent by two
different sites between successive returns would in general be
different for systems without periodic external forcing, and this was
precisely the motivation for the asynchronous update rule in
\cite{Tomas}.  However, here we wish to mimic the variations in return
times in a {\em deterministic } fashion. This motivated us to
introduce a second field in the CML. Note that the simplest chaotic
oscillator would be a system of three phase space dimensions, like the
Lorenz equations. Applying a Poincar\'e section reduces such a system
to a two-dimensional map.  This is also the case in systems with
external periodic forcing. Here the simplest realization would be
systems like a damped nonlinear pendulum or Duffing oscillator with a
time-periodic forcing. A Poincar\'e section again reduces the system
to a two-dimensional map and after the synchronous iteration the
respective units are still at equal time.

We therefore replace the single variable map $f(u)$ used in Eq.
(\ref{cm}), by a new map with an additional variable $v$:
\begin{eqnarray}
  u_i(n+1)&=&f(u_i(n) +\frac{\varepsilon}{2}\Delta_fu_i(n)+v_i(n)~,
\label{2dima} \\
v_i(n+1)&=&b(u_i(n+1)-u_i(n))~.\label{2dimb}
\end{eqnarray}
Here $f$ is the same map as before and the new parameter $b$ is the
Jacobian of the full two-dimensional local map; this map is invertible
for any non-zero $b$ and becomes increasingly two-dimensional with
$|b|$.  The change in the local map (\ref{cm}) is analogous to how the
2D H\'enon map ~\cite{henon} is constructed from the 1D logistic map,
except that $b(u_i(n+1)-u_i(n))$ appears here on the right hand side
instead of $bu_i(n)$. This ensures that the absorbing state fixed
points $u_i(n)=u^*$ of the old CML (\ref{cm}) are mapped to the
laminar fixed point $(u_i(n),v_i(n)) = (u^*,0)$ in the new CML. The
model (\ref{2dima}-\ref{2dimb}) is a completely deterministic system
with no element of stochasticity and is updated synchronously.  The
value of $u_i$ determines, as in model (\ref{cm}), whether a given
site is ''active'' or ''inactive''. Starting from the
Chat\'e-Manneville case ($b=0$) we can follow the transition between
laminar and chaotic states. As we will see below, the new parameter
$b$ actually opens up the possibility to study the effect of the
``solitons'' on the dynamical states and transitions of CMLs; this
appears to be a more important issue than the dimensionality of the
local map.

\subsection{Qualitative properties.}\label{sssol}

\begin{figure}[t]
\includegraphics[width=8.5cm]{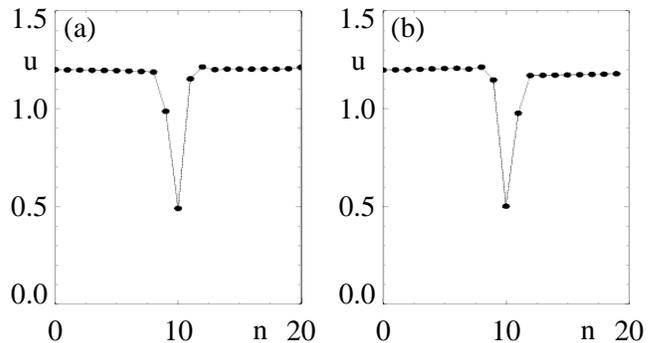}
\caption[]{Average profiles of a right (a) and left (b) moving soliton
that occurs in the Chat\'e-Manneville model at criticality for $r=3$.
}\label{soliton}
\end{figure}

\begin{figure}[t]
\includegraphics[width=8.5cm]{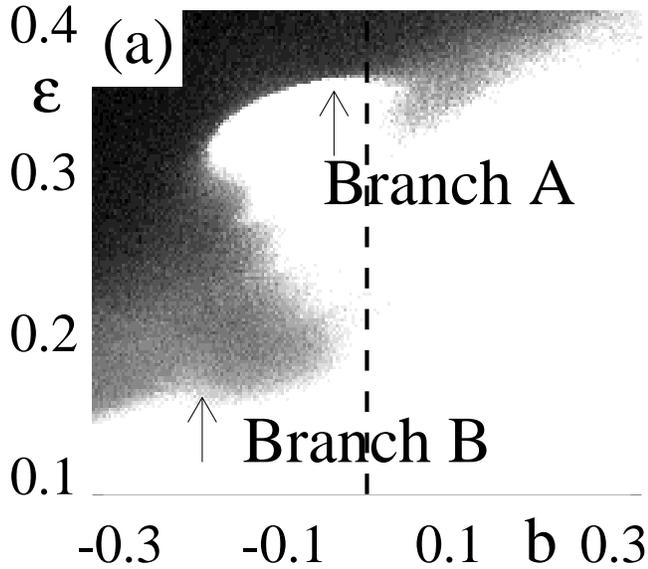}
\caption{ Activity in the model (\ref{2dima})-(\ref{2dimb}) at
$t\!=\!1000$.  White regions correspond to points in the
$(\varepsilon,b)$-plane where the initial activity has decayed into an
absorbing configuration and the darker regions to points with a
non-vanishing order parameter. Clearly the transition curve becomes
quite complicated; the two branches discussed in this paper are
indicated as ``Branch A'' and ``Branch B'' (see text).  The dashed
line indicates the Chat\'e-Manneville model ($b\!=\!0$) }\label{fig1}
\end{figure}

\begin{figure}[h]
\includegraphics[width=8.5cm]{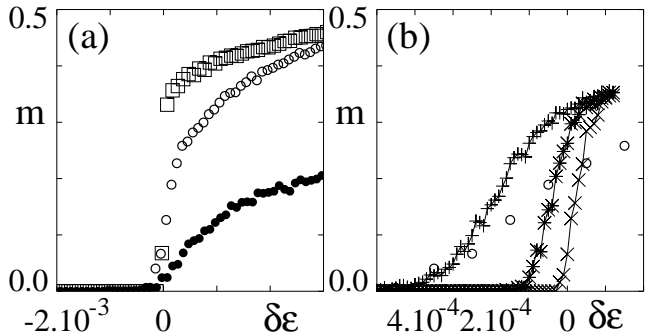}
\caption[]{ Activity in the model (\ref{2dima}-\ref{2dimb}) for
128 systems of size $L=2048$ and $r=3$. (a) Activity as a function of
$\delta \varepsilon$ (distance to the critical point) at time $2
\times 10^5$ for $b=-0.1$ (squares), $b=0$ (open circles) and
$b=0.2$ (closed circles).  The transition appears much sharper for
negative values of $b$.  (b) Steepening for the transition at
$b=-0.1$ for increasing times: $5 \times 10^3$ (+), $5 \times
10^4$ (*), $5 \times 10^5$ ($\times$). To stress the magnified
scale of $\delta \varepsilon$, also the data shown in panel (a)
for $b=0$ is plotted (open circles). }\label{fig2}
\end{figure}

Our CML now contains three freely adjustable parameters ($r,
\varepsilon$ and $b$), and clearly we will have to focus on a
subset of parameters. Our main focus will be on the case where
$r=3$, although we will also study the transition for $r=2.2$. For
$r=3$ and $b=0$ the dynamics shows many solitons (see
Fig.~\ref{soliton}) and the critical exponents appear to differ
significantly from those of DP.

To get a feeling for the location of the transition as function of
$b$ and $\varepsilon$, we show in figure \ref{fig1} the activity
(defined as the average number of active sites) after 1000
iterations in the ranges $-0.3 \leq b \leq 0.3$ and $0 \leq
\varepsilon \leq 0.4$ at $r\!=\!3.0$. The ``traditional''
transition is that occurring at $b\!=\!0$ and $\varepsilon\!=\!
0.35984\dots$. Clearly, for negative values of $b$ two additional
transitions emerge. Here we only study points on the two
transition branches labeled ``A'' and ``B'' in Fig.~\ref{fig1};
below we focus on the behavior along branch A.

\subsubsection{Qualitative changes in behavior along branch A}

\begin{figure}[t]
\includegraphics[width=8.5cm]{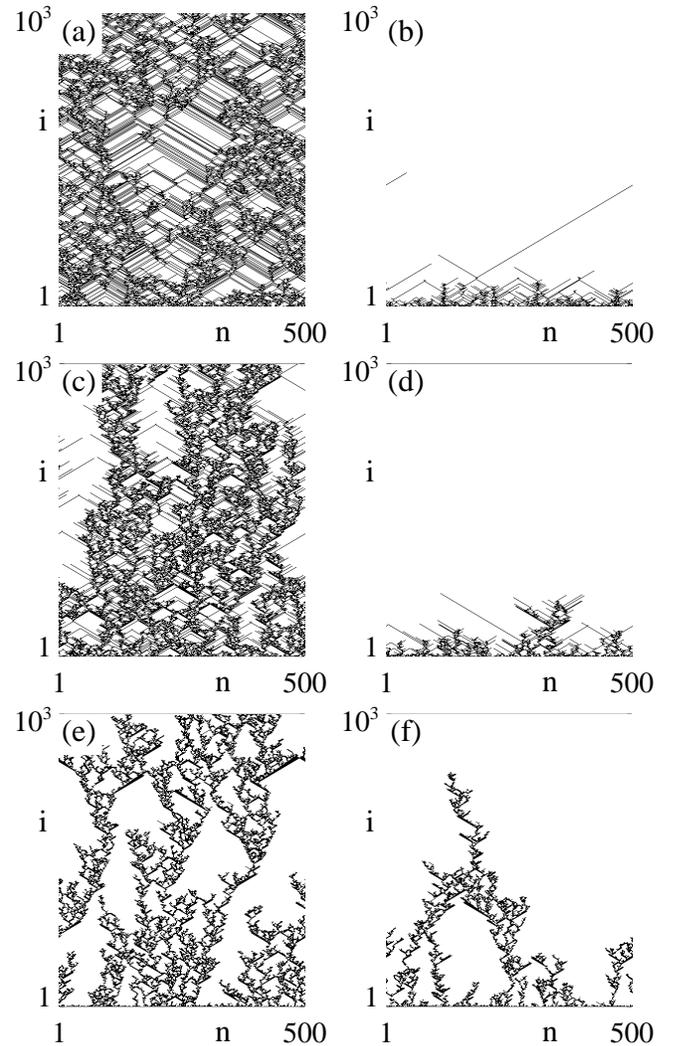}
\caption[]{Spacetime plots of our coupled map lattice
(\ref{2dima}) -(\ref{2dimb}) for $r\!=\!3$ above (left column) and
below (right column) criticality. Inactive sites are white,
chaotic sites are black. (a) $b=-0.1$, $\varepsilon=0.353$ (b)
$b=-0.1$, $\varepsilon = 0.343$ (c) $b=0$ $\varepsilon= 0.361$ (d)
$b=0$, $\varepsilon = 0.351$ (e) $b=0.2$, $\varepsilon = 0.374$
(f) $b=0.2$, $\varepsilon = 0.364$. }\label{fig3x2}
\end{figure}

Figure \ref{fig1} hints that the sharpness of the transition varies
along branch A: the jump in order parameter appears to become steeper
for negative values of $b$. The differences in the nature of the
transitions are illustrated more clearly in Fig.~\ref{fig2} by
plotting the value of the order parameter as a function of
$\varepsilon$ for $b=0$ and $b=-0.1$ averaged over an ensemble of 32
systems for a number of times. The behavior for $b=0$ is consistent
with a continuous transition, whereas for $b=-0.1$ longer times lead
to a marked steepening, consistent with the emergence of a
discontinuity.

\paragraph{Soliton regime}

Some effects of the parameter $b$ on the dynamics can also be seen
from the evolution of the binary patterns at $r=3$
(Fig.~\ref{fig3x2}). For $b=0$, solitons can be seen both above and
below threshold (Fig.~\ref{fig3x2}c-d). They consist of pairs of
active sites and propagate with velocity one. Their maximal lifetimes
are of order 100 (Fig.~\ref{fig3x2}d). When $b$ is decreased to a
value of $-0.1$, the typical lifetimes of solitons become so long that
they typically only vanish when they collide with other solitons or
propagate into turbulent structures. When two solitons collide, they
either annihilate or create new turbulent structures. Such creation is
clearly visible in Fig.~\ref{fig3x2}a for $n\approx 200$ and $i\approx
600$.

For sufficiently large $b$ the isolated solitons present in the
original model $(b=0)$ are suppressed: solitons with a lifetime
longer than a few iterations are rare here. On the other hand,
there are regular ``edge'' states visible, where an active state
propagates ballistically while emitting new activity; one example
is visible in Fig.~\ref{fig3x2}e for $n\approx400$ and
$i\approx800$. These structures do not seem to influence the order
of the transition, but they may very well lead to rather large
crossover scales.

\begin{figure}[t]
\includegraphics[width=8.5cm]{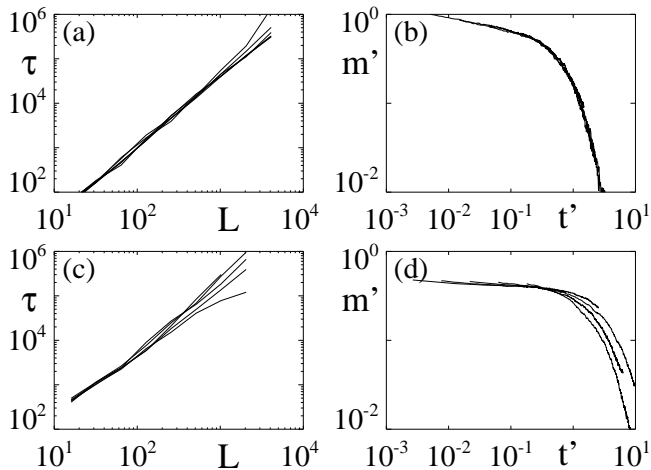}
\caption[]{Examples of good rescaling plots for $b=0.2$ (a,b) and poor
rescaling for $b=-0.1$ (c,d). (a) Absorption time $\tau$
vs. systemsize L, for $r=3, b=0.2$ and
$\varepsilon=0.3727,0.37322,0.37323$ (critical value), $0.37325,
0.3733$ and $0.3735$. (b) Rescaled average activity $m':=m
L^{-\theta}$ versus rescaled time $t':=t/L^z$ for $r=3, b=0.2$ and
$\varepsilon=0.37323$ for $L=32,64,128,256$ and $512$, showing a good
data collapse.  (c) Absorption time $\tau$ vs. systemsize L, for $r=3,
b=-0.1$ and $\varepsilon=0.35200,0.35203,0.35205, 0.35206$ and
$0.35207$.  Even small changes in $\varepsilon$ lead to substantial
changes in the absorption time, and it is difficult to estimate the
critical value of $\varepsilon$. (d) Rescaled average activity $m'$
versus rescaled time $t'$ for $r=3, b=-0.1$ and $\varepsilon=0.35203$
for $L=64,128,256,512$ and $1024$, showing poor data collapse; either
the initial decay or the tails do not overlap; shown here is a
compromise. Note that the initial decay is very slow, leading to a
small estimate for the value of $\theta$.}\label{figa39a34}
\end{figure}

In conclusion: the value of $b$ has a large influence on the presence
of solitons, and also influences the steepness of the transition. In
fact; discontinuities are found at points in $(\varepsilon,b)$-space
where solitons dominate the dynamics. This implies that the
(colliding) solitons have a strong influence on the global dynamics
and are able to change the nature of the transition from a continuous
to what appears as a first order one.  We will make this point more
precise below.

\subsection{Finite size scaling in 2nd order regime}\label{ssfinite}

Stochastic systems belonging to the DP universality class are
characterized by a set of critical exponents describing, e.g., the
order parameter $m(\varepsilon,L,t)$ and the behavior of the
``absorption time'' $\tau(r,\varepsilon,L)$, i.\ e.\ the averaged time
it takes the system, starting from a random initial state, to reach
the absorbing state. From finite-size scaling arguments
~\cite{Mogens}, one finds that the order parameter $m$ at the critical
point $\varepsilon_c$ should behave as
\begin{equation}
m(L,t) \sim L^{-\beta/\nu_{\perp}} g(t/L^z)~. 
\end{equation}
For a finite lattice, the absorption time $\tau$ then increases as
\begin{equation}
\tau \sim L^z~.
\end{equation}
Finally, for short times ($t \ll L^z$), $g(t/L^z) \sim
(t/L^z)^{-\beta/\nu_{\parallel}}$, so that for short times
$m$ should decay as
\begin{equation}
 m(L,t) \sim t^{\theta} ~~ for~~ t<<L^z
\end{equation} 
Here the usual dynamical exponent $z=\nu_\parallel/\nu_\perp$ has been
introduced, defined as the ratio between the correlation length
exponent in the time direction $\nu_\parallel$ and the correlation
length exponent in space $\nu_\perp$. The scaling relation $\theta =
{-\beta/\nu_\parallel}$ connects the critical exponents.

\begin{table}[t]
\begin{ruledtabular}
\begin{tabular}{|c|l|l|l|l|}
$r$ & $b$   & $\varepsilon_c$ & $z$     & $\theta$ \\ \hline
 2.2 & -0.02 & 0.01338()    & 1.57(1) & 0.16(1) \\
     & -0.01 & 0.01465()    & 1.57(1) & 0.16(1) \\
     & ~0.0  & 0.01605(2)   & 1.53(1) & 0.17(1) \\
     & ~0.01 & 0.017628(5)  & 1.57(1) & 0.16(1) \\
     & ~0.02 & 0.01921(2)   & 1.57(2) & 0.17(2) \\ \hline
 3.0 & -0.25 & 0.16312(3)   & 1.58(1) & 0.160(5)\\
     & -0.2  & 0.16495(2)   & 1.58(2) & 0.168(3)\\
     & -0.15 & 0.16205(1)   & 1.58(1) & 0.17(1) \\
     & -0.125& 0.16368(2)   & 1.57(1) & 0.20(1) \\
     & -0.1  & 0.35203(1)   & 1.52(3) & 0.02(2) \\
     & ~0.0  & 0.35984(3)   & 1.42(2) & 0.18(1) \\
     & ~0.1  & 0.3393(1)    & 1.48(2) & 0.155(1) \\
     & ~0.125& 0.34745(5)   & 1.53(2) & 0.15(1) \\
     & ~0.15 & 0.35680(5)   & 1.57(1) & 0.159(3)\\
     & ~0.175& 0.36545(1)   & 1.58(1) & 0.16(1) \\
     & ~0.2  & 0.37323(1)   & 1.58(1) & 0.16(1) \\ \hline
DP  &       &              & 1.58074 & 0.15947
 \end{tabular}
\end{ruledtabular}
\caption{\label{tab:table1} The critical exponents $z$ and
$\theta=\beta/\nu_\parallel$ for our CML. The values for DP are
taken from ~\cite{DP-val}. }
\end{table}

\begin{figure*} [t]
\includegraphics[width=16cm]{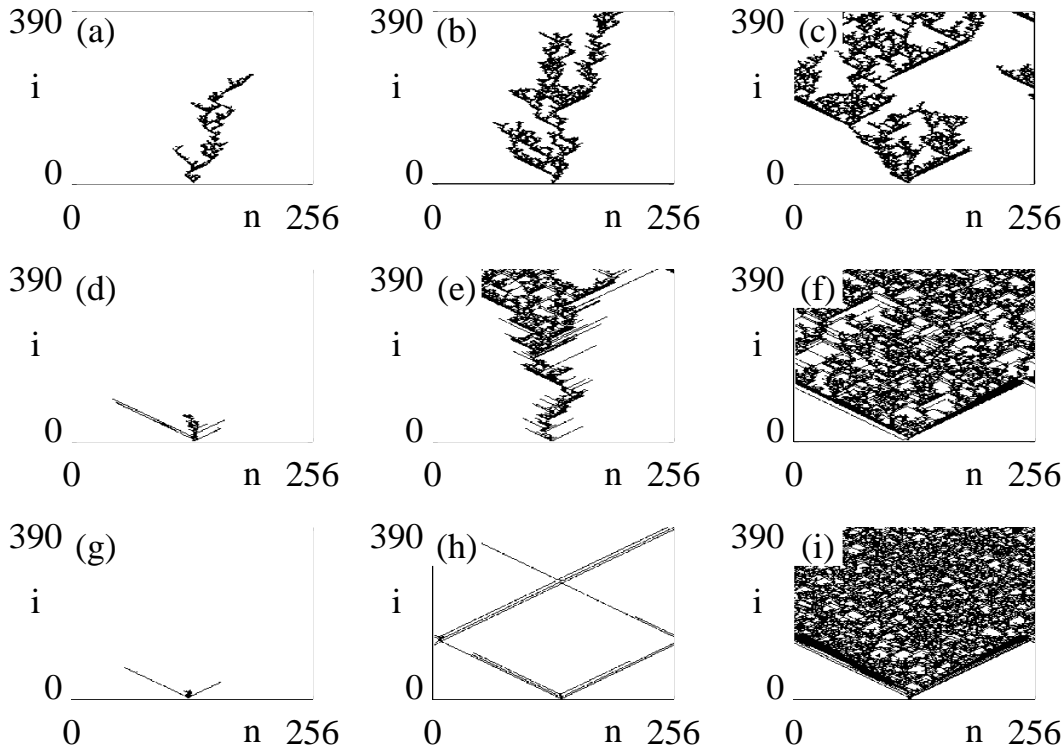}
\caption{\protect \small{ {\sf Spreading of a single turbulent seed
through the ``natural-initial-state'' below criticality, at
criticality and above criticality, for $b=0.2$ (a-c), the
Chat\'e--Manneville model for $b\!=\!0$ (d-f) and $b=-0.1$ (g-i).  In
all cases $r=3.0$. }}} \label{spr727}
\end{figure*}

To estimate the critical exponents for our CML we performed direct
numerical simulations and calculated the absorption time $\tau$ and
the order parameter $m$, defined as the average activity.  We used
ensembles of initial conditions, in which all initial $u$ values are
assigned a random value in the chaotic phase, $0\leq u_i(0) \leq 1$.
The $v$-values of the initial state are set to zero to ensure that
they do not influence the $u$-values from the onset of iteration and
that the analogy with the original model and our variant at $b=0$ is
satisfied.

The behavior of the absorption time at criticality is used to
determine the critical point and the $z$ exponent \cite{Tomas,Mogens}.
An ensemble of $128$ systems is iterated forward in time until an
absorbing configuration is reached. The average number of time steps
needed before reaching such a configuration yields the absorption time
$\tau$. Examples of $\tau$ as function of $L$ are shown in
Fig.~\ref{figa39a34}a,c; the best fit to a straight slope determines
the critical exponent $z$.

In Fig.~\ref{figa39a34}b,d we plot examples of $m':= m L^{-\theta}$ as
function of $t':= t/L^z$ for a range of $L$'s. When proper scaling
occurs, as is the case in Fig.~\ref{figa39a34}b, the curves for
different $L$ fall on top of each-other, and the initial power-law decay
of $m'$ determines the exponent $\theta$.  Here an ensemble of $1000$
systems was used. The order parameter was calculated as the sum of
active sites divided by the total amount of sites.  The systems are
iterated forward until $t \approx L^z$, where the algebraic behavior
clearly ends.

Estimates of critical exponents have been done for $r=2.2$ and
$r=3.0$.  For $r=3.0$, the critical exponents for the original model
($b=0$) shows significant deviations from the corresponding DP values
and the computational costs are tolerable. In Fig.~\ref{figa39a34}a,b
we show examples of the rescaling plots for $r=3,b=0.2$, where a nice
data-collapse occurs and the transition appears to be of 2nd order,
and for $r=3,b=-0.1$ (Fig.~\ref{figa39a34}c,d), where the
data-collapse is poor and the transition appears to be no longer
continuous.

The values of the critical exponents are given in Table
\ref{tab:table1} and correspond simply to the best possible values,
irrespective of the quality of the data collapse. For $r=2.2$ DP
values are found for $|b| \geq 0.01$. For $r=3$, the critical
transition on Branch B appear to be DP-like, while on Branch A a
crossover to DP values is found when $b$ is large enough ( $|b|>
0.15$). This regime coincides with values of $b$ where the solitons
are suppressed in the space-time plots, and a continuous transition
takes place. The soliton dominated dynamics at $b=-0.1$ is reflected
in the extremely low value of the exponent $\theta$, characterizing
the decay of the order parameter. Here the data-collapse is rather
poor as shown in Fig.~\ref{figa39a34}c,d.


\begin{figure*}[t]
\includegraphics[width=16cm]{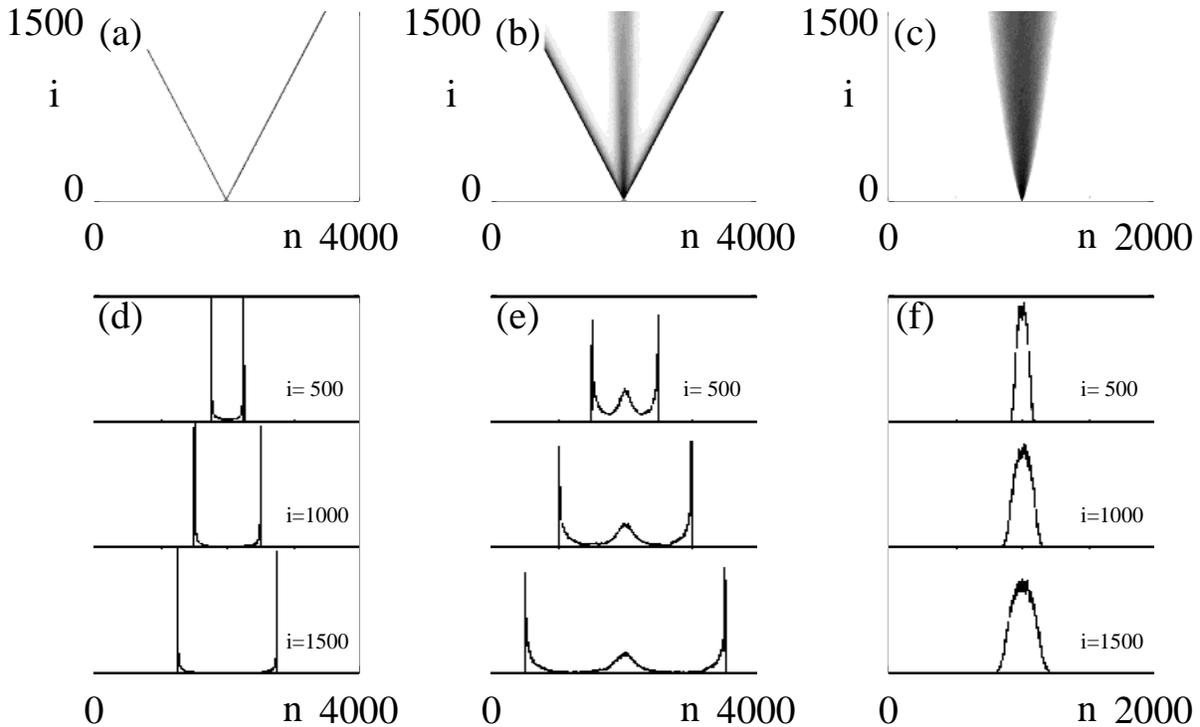}
\caption[]{The average spreading of active seeds in the
``natural-initial-state'' close to criticality. The densities are
obtained by averaging over $10^4$ realizations, and for clarity we
have included three snapshots of this average activity in the bottom
rows of this figure. Parameters are $r=3$ (for all runs), and $b=-0.1$
(a,d), $b=0.0$ (b,e) and $b=0.2$ (c,f). The respective values of
natural-initial-state are $1.170$, $1.212$ and
$1.235$. }\label{figspread}
\end{figure*}

\subsection{Spreading of Turbulence in 2nd order regime.}\label{ssspread}

So far the critical properties of the CML's starting from
``homogeneous'' states have been studied, i.e., with initial
conditions where each site in the lattice is assigned a random number
in the chaotic (turbulent) phase.  A different approach is to consider
the spreading of a single turbulent seed in an otherwise laminar
configuration (see Fig.~\ref{spr727}).  This makes it possible to
study the dynamical critical exponents, or spreading exponents, and
see how these compare to the directed percolation counterparts.

For spreading of activity in stochastic systems with absorbing states
the following quantities are characterized by critical exponents
\cite{grasstor}: the total number of chaotic sites $N(t)$, the
survival probability $P(t)$, the mean-squared deviation $R^2(t)$ of
the turbulent activity from the ``seed'' and the density $n(t)$ of
chaotic sites within the spreading patch of turbulence.  It is assumed
that they behave according to
\begin{equation} N(t) \sim t^{\eta_s} ~~ P(t) \sim t^{-\delta} ~~
R^2(t) \sim t^{z_s} ~~ n(t) \sim t^{-\theta_s} \end{equation} For
probabilistic systems it has been conjectured and verified numerically
~\cite{Mendes,hyperscaling}, that the dynamical exponents satisfy the
generalized hyper-scaling relation $\eta_s + \delta + \theta_s = d
z_s/2$ where $d$ is the spatial dimension.  For systems with a single
absorbing state, including DP, one finds that $\delta = \theta_s =
\beta/\nu_\parallel$ and $z_s=2/z$, reducing the hyper-scaling
relation to $4\delta +2\eta_s = d z_s$.

Systems with infinite numbers of absorbing states have been studied
carefully recently and it has been found that they differ from the
classical ones (with a single absorbing state belonging to the DP
universality class) precisely in the non-universality of the spreading
exponents \cite{hyperscaling}.  Only exponents characterizing
quantities averaged over surviving runs alone are found to be
universal.  This implies that $z_s$, the sum $\eta_s + \delta$ and
$\theta_s$ are expected to be universal, while $\eta_s$ and $\delta$
individually are not.  Only for the so called
``natural-initial-state'' are the DP values found for the exponents
characterizing quantities averaged over all runs. Such a particular
state is constructed by letting the system evolve at criticality from
homogeneous initial conditions, where all sites initially are in the
active phase, until an absorbing configuration is reached.

After a few spreading experiments in our CML we indeed observed
that the propagation of activity from the initial seed through the
laminar region depended strongly on the configuration of the
laminar state surrounding the seed.  Moreover the dynamical
exponents varied with this configuration, thus being
non-universal. So the non-unique absorbing state of our CML (any
configuration with all $u$-values above unity and $v$-values not
too large will be absorbing) leads to behavior as can be expected
for DP with an infinite numbers of absorbing states.

We determined the ``natural-initial-state'' by iterating systems
of up to $4096$ sites from homogeneous initial conditions until an
absorbing configuration is reached. The average value of all sites
is then used as the value of the laminar background.  For $r=3.0$
we have calculated these as $1.235$ for $b=0.2$, $1.212$ for
$b=0$, $1.170$ for $b=-0.1$ and $1.0395$ for $b=-0.2$.

 \begin{figure}[t]
\includegraphics[width=8.5cm]{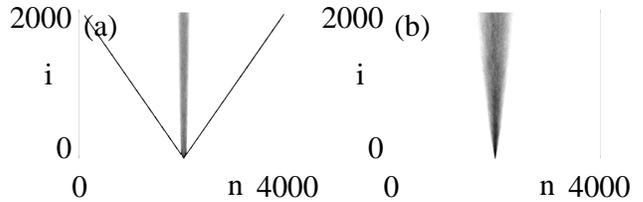}
\caption[]{Average spreading activities near criticality for
$r=3,b=0.2$ (a) and $r=3,b=0$ (b). In comparison to the spreading into
the ``natural-initial state'' as shown in Fig.~\ref{spr727}, the value
of $u$ in the laminar background has been lowered from 1.235 to 1.225
in (a), thus strongly enhancing solitons, and has been increased from
1.212 to 1.22 in (b), thus strongly suppressing soliton activity.
}\label{figspreadalter}
\end{figure}

In Fig.~\ref{figspread} we display the average spreading for $r=3$ and
$b=-0.1, 0$ and $0.2$. Clearly, for $b=-0.1$ and for the
Chat\'e--Manneville model at $r=3,b=0$ it is basically impossible to
estimate the spreading exponents at the ``natural-initial-state'',
since the spreading is dominated by the solitons (see
Fig.~\ref{figspread}b,e). This behavior is distinctly different from
what is observed in the various systems belonging to the DP
universality class. We have therefore only estimated the spreading
exponents for $b=0.2$ at branch A, and $b=-0.2$ and branch B; in both
cases the solitons are not dominant.

Note that the strength of the spreading solitons can be altered by
changing the value of the laminar background.  By increasing the
background to a value above the ``natural-initial-state'' one, the
solitons can be suppressed, while they can be enhanced by a decrease
of the laminar value; see Fig.~\ref{figspreadalter}.

 \begin{table}[t]
\begin{ruledtabular}
 \begin{tabular}{|c|l|l|l|l|l|l|}
 $b$ &$x_i$ & $z_s $& $\delta$ & $\eta_s$& $\theta_s$& $\Delta$ \\
\hline
 0.2&1.229 & 1.98(2)& 0.00(0) & & & \\
 &1.23 & 1.68(2)& 0.10(1)&0.43(2) & 0.32(1) & -0.01(2)\\
 &1.235 & 1.60(1)& 0.16(1)& 0.34(1) & 0.29(1) & 0.01(1) \\
 &1.24 & 1.61(2)& 0.23(2)& 0.25(1) & 0.31(1) & 0.01(2) \\
 &1.245 & 1.65(1)& 0.30(1)& 0.23(1) & 0.29(2) & 0.00(1) \\
 &1.25 & 1.65(3)& 0.34(2)& 0.14(1) & 0.300(3)& 0.05(2) \\
 &1.255 & 1.69(2)& 0.35(2)& 0.14(2) & 0.29(1) & 0.07(2) \\
 &1.26 & 1.72(3)& 0.43(1)& 0.04(1) & 0.29(1) & 0.10(2) \\ \hline
 -0.2&1.13       & 1.99(1) & 0.00(0) & 0.793(2)& 0.205(1)& 0.00(1) \\
 &1.135  & 1.59(2) & 0.09(1) & 0.42(1) & 0.29(1) & -0.01(1)\\
 &1.139  & 1.58(2) & 0.170(3)& 0.32(1) & 0.30(1) & 0.00(1) \\
 &1.1395  & 1.58(1) & 0.16(1) & 0.31(1) & 0.28(1) & 0.04(1) \\
 &1.145  & 1.56(1) & 0.249(2)& 0.20(1) & 0.24(2) & 0.08(2) \\
 &1.15   & 1.61(1) & 0.347(3)& 0.11(1) & 0.27(1) & 0.08(1) \\
 &1.155   & 1.68(3) & 0.45(1) & 0.01(1) & 0.22(1) & 0.16(1) \\ \hline
DP & & 1.26523&0.15947&0.31368&0.15947& \\
 \end{tabular}
\end{ruledtabular}
\caption{\label{tab:table2}Estimated spreading exponents for $r=3.0$
for background values $x_i$. The deviation from the hyper-scaling
relation for $d=1$ is defined as $\Delta \equiv
z_s/2-(\eta_s+\delta+\theta_s)$. Note that for $b=0.2$, the natural
background state has $x_i \approx 1.235$; for this value, the
exponents $\delta$ and $\eta_s$ are close to their DP values.
Similarly for $b=-0.2$, the natural background state has $x_i \approx
1.1395$; again $\delta$ and $\eta_s$ are close to their DP values}
\end{table}

Our estimates of the dynamical exponents have been done for
simulations with a maximum time of 2000 iterations. An active seed is
placed in the center of the lattice, surrounded by a laminar
background. The seed consists of two active sites, each of which is
assigned a random number in the chaotic regime, such that the location
is fixed but the values of the active sites differ for each trial in
the ensemble.  The ensemble size $N_s$ used for statistical averaging
and the number of sites in the lattice $L$ have been adjusted to the
number of surviving runs the different setups produced, and how far
the turbulence propagated out from the seed. A minimum of 200
surviving runs have been used in the averaging.

Our results in table \ref{tab:table2} agree rather well with
previously obtained results for probabilistic systems with an infinite
number of absorbing states. In particular, the exponents averaged over
surviving runs alone definitely seem to be universal as long as the
background does not deviate to much from the ``natural-initial-state''
one.  While the values for the sum $\delta + \eta_s$ are very close to
the DP value of $0.47315(7)$, our results for $z_s$ and $\theta_s$
deviate from their respective DP values.  A very interesting
observation is that the hyper-scaling relation is satisfied ($\Delta
\simeq 0$) for the majority of different background values. Only for
the highest values are significant deviations encountered.

\section{Stochastic Model.}\label{sstoch}

The propagating structures, which are observed in the CML that we
studied in the previous section, appear to play an important role for
the transitional behavior. It is, however, numerically very demanding
to obtain good statistics for large CMLs and long times. As pointed
out already in the introduction, this is the reason why one tries to
map such deterministic models to simple stochastic models. Not only
may there be more hope to understand such models analytically, they
also are much easier to handle from a computational point of view.

In this section we will introduce and study a very simple extension of
the Domany-Kinzel cellular automaton which itself is a simple model
showing DP behavior. While for the Domany-Kinzel automaton every site
can only be active or inactive, we will allow sites to either contain
a left or right traveling soliton. As in the CML, these solitons
should be generated from active sites only, and we wish to be able to
tune their typical lifetime. The only process in which these solitons
aid the spreading of activity is by collisions: for simplicity we
assume that with probability {\em one} a pair of colliding solitons
yields a singe active site.

Below we will first discuss the definition of our model in section
\ref{ssdef}. We will then discuss the mean field equations for our
model in section \ref{ssmft}, and these will show a transition
from second to first order behavior. We will study the statistical
properties of our model  in section \ref{ssphe}. We will
 illustrate the role
of solitons in direct simulations of this model; these simulations
will point to the relevance of large ''holes'' which cannot be
``healed'' by the solitons.  We will discuss the statistical
properties of our model near the transition from inactive to
active states in the soliton-dominated parameter regime. We will
find that the transition is no longer in the DP universality
class, since no asymptotic scaling regime can be reached.  While
the transition shows some characteristics of a first order
transition (dependence on initial state for example), the
asymptotic situation is not entirely clear: rather we find a
regime of long lived transient states between active and inactive
regimes.

\begin{figure}[t]
\includegraphics[width=8.5cm]{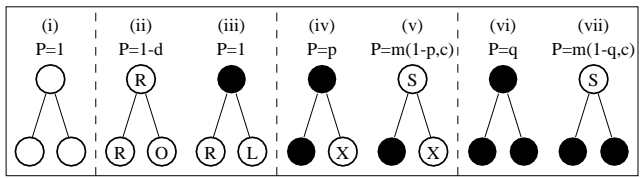}
\caption[]{Schematic definition of our stochastic model. In this
picture, the bottom two circles denote possible incoming states,
the circle at top denotes the possible stochastic outcome. Our
model is defined on a diamond lattice and so one only needs to
define the probabilities for certain offspring (active, inactive,
right or left traveling soliton) as a function of its two
predecessors. Empty circles depict inactive states, black circles
are active, ``R'' and ``L'' denote right and left-moving solitons,
``S'' denotes a soliton of arbitrary direction, and ``X'' finally
represents any state (L,R,inactive or active). The notation
$m(a,b)$ denotes the minimal value of $a$ and $b$, and $P$ denotes
the conditional probability that this out-coming state occurs.
 See text for a more elaborate explanation.}\label{stoch}
\end{figure}

\subsection{Definition of model}\label{ssdef}
The (1+1) dimensional Domany-Kinzel cellular automaton is defined on a
diagonal square lattice where each site can either be active or
inactive. The model evolves by parallel updates according to the
transition probabilities $p$ and $q$, corresponding to the
probabilities that an empty plus an active site or two active sites
respectively produce a single active site.  The choice $q=p(2-p)$
corresponds to a realization of directed bond percolation \cite{DK}.

In our extension the active sites behave like usual directed bond
percolation except from the fact that with probability $c$ they can
emit a left or right-moving soliton.  These solitons have a tunable
lifetime and travel ballistically. We assume that the solitons cannot,
by themselves, create chaos, except when two solitons collide.

The updating rules are illustrated in figure \ref{stoch}, where the
sites can be either inactive (empty), active (black), or contain a
left (L) or a right (R) moving soliton (S).

\noindent{\bf (i)} The inactive state: two inactive sites always yield
an inactive site. This property ensures that there is a unique
absorbing state.

\noindent{\bf (ii)} Soliton propagation: a right moving soliton (R)
either dies with probability $d$, or propagates with probability
$(1-d)$ when the ``O' state to its right is inactive or another
right-moving soliton. The rule for left-moving solitons follows by
left-right symmetry.

\noindent{\bf (iii)} Soliton collision: when two oppositely
propagating solitons collide, they generate an active site with
probability one.  This is the only process where solitons lead to
spread of active sites. In principle we could generate active
sites with a probability less than one, but it may be expected
that this does not change the behavior of the model in a
qualitative sense.

\noindent{\bf (iv)} Single active sites: a single active site, where X
can either be a soliton or inactive site, leads with probability $p$
to a new active site. Note that the spreading of activity is thus not
enhanced by individual solitons

\noindent{\bf (v)} Transformation: a single active site can give rise
to a soliton (S) with probability $min(1-p,c)$; $c$ denotes the
creation rate of solitons. Such a new soliton can be either left or
right-moving with equal probability.

\noindent{\bf (vi)} Pair of active sites: two active sites create a new
particle with probability $q$; we restrict ourselves to bond-directed
percolation and take $q=p(2-p)$.

\noindent{\bf (vii)} Soliton creation from pair of active states:
similar to case (v), a pair of active sites can give rise to solitons
with probability $min(1-q,c)$.

\begin{figure*}
\includegraphics[width=16cm]{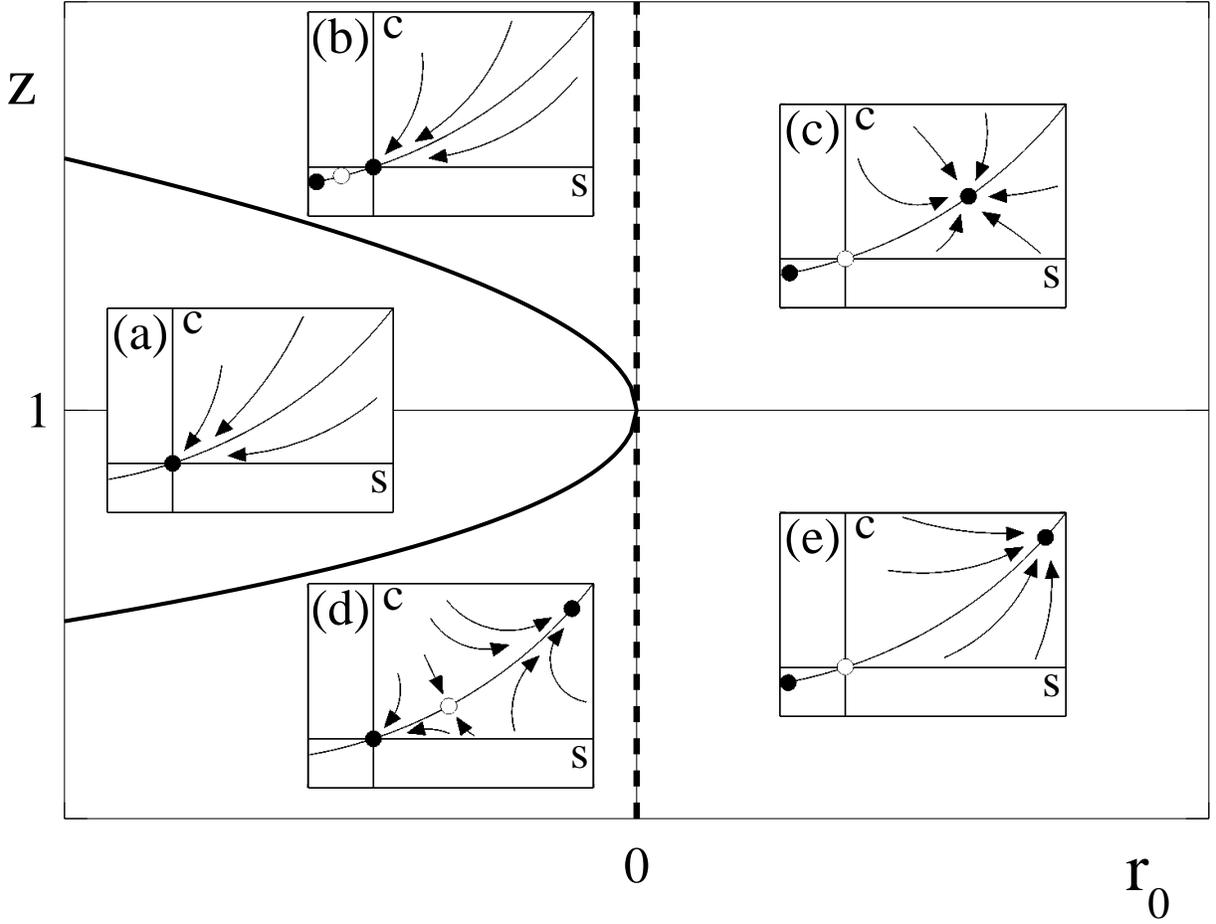}
\caption[]{Dynamical system analysis of the mean field equations
  (\ref{resc_s})-(\ref{resc_c}). The full and dashed curves show the
  location in $r,z$ space of the saddle-node and transcritical
  bifurcations respectively. The five insets (a-e) show schematically
  the flow in the various regimes of the mean field equations. For
  more details see text.}\label{figflow}
\end{figure*}

\subsection{Mean Field Equations.}\label{ssmft}

To interpret the physical properties of our cellular automaton, a
crude insight can be obtained by applying Mean-Field Theory.  In this
approximation it is reasonable to ignore the differences between left
and right traveling solitons, and so our mean field equations are for
two concentrations, those of chaotic sites $c$ and solitons $s$.

\paragraph{Equation for Chaotic Sites:}
Chaotic sites can emit solitons and can be generated by collisions of
two solitons; apart from these two rules they behave like DP. Thus,
without the solitons the rate equation (without noise) would be
$\dot{c} = b_1 c - b_3 c^2$ \cite{DP2}. To incorporate the creation of an
active site when two solitons collide according to rule (iii), the
term $b_2 s^2$ needs to be added to this equation. There is no source
term linear in $s$ in the rate equation for $c$, reflecting that we
assume that individual solitons do not give rise to activity.

\paragraph {Equation for Solitons}: There are four processes that influence
the solitons. Solitons may decay spontaneously according to rule (ii),
and this yields a term $- a_3 s$ in the rate equation for $s$.
Solitons also die upon collision leading to a term $\propto - s^2$.
Depending on the lifetime of the solitons, either of these two terms
may dominate and so we keep both of them; we will see below that this
will indeed be a crucial ingredient. Solitons are created from active
sites according to rule (v) and (vii). While this in principle yields
source terms in the rate equation of $s$ proportional to both $c$ and
$c^2$, we only keep the linear term, since the prefactor for both
these terms will be of the same order.  Inclusion of the quadratic
term does not affect the qualitative dynamics.

The rate equation for the solitons and chaotic sites can then be
written as
\begin{eqnarray}
\dot{s}&=&a_1c-a_2s^2-a_3s \\
\dot{c}&=&b_1c+b_2s^2-b_3c^2
\end{eqnarray}
where the lifetime of the solitons is set by $1/a_3$ and the spreading
rate of the chaotic patches by $b_1$.

These two equations can be simplified by the introduction of a
rescaled of time $\tau$ and densities $S$ and $C$ to be
\begin{eqnarray}
  \dot{S}&=&C-S^2-aS \label{resc_s} \\ \dot{C}&=& r_0 C+S^2-uC^2
\label{resc_c}
\end{eqnarray}
where
\begin{eqnarray}
a=\frac{a_2 a_3}{a_1 b_2}, &~~ r_0=\frac{a_2 b_1}{a_1 b_2}, &~~
u=\frac{b_2 b_3}{a^2_2} \label{rr1}\\ 
t=\frac{a_2}{a_1 b_2} \tau, &~~
s=\frac{a_1 b_2}{a^2_2}S, &~~ c =\frac{a_1 b^2_2}{a^3_2} C \label{rr2}
\end{eqnarray}

We will now analyze the possible transitions in the mean field Eqs.
(\ref{rr1}-\ref{rr2}).
\paragraph{Fixed points }
The fixed points $(S^*,C^*)$ of the rescaled equations
(\ref{resc_s})-(\ref{resc_c}) satisfy $C^*=S^{*2}+aS^*$, where $S^*$
is given by solutions to the fixed point equation
\begin{eqnarray}
Sf(S)&=&0 \\
f(S)&=&uS^3+2uaS^2+(ua^2-1-r)S-ar_0 \label{reduced}
\end{eqnarray}
Apart from the trivial fixed point $(S^*,C^*)=(0,0)$ there may be
either 1 or 3 other fixed points which can be found from solving Eq.
(\ref{reduced}).  It can be shown that Eq. (\ref{reduced}) always has
one solution for large negative $S$.  This fixed point can be ignored
since only points where both $S^*$ and $C^*$ are positive are relevant
for our mean field equations (remember that $S$ and $C$ are both
concentrations).  The two non-trivial fixed points $(S_1^*,C_1^*)$ and
$(S_2^*,C_2^*)$ are born in a Saddle-Node bifurcation when the
discriminant of Eq. (\ref{reduced}) becomes negative. Introducing the
parameter $z:= a^2 u$ and performing the tedious standard algebra
yields that this occurs when
\begin{equation}
z^2 - (2 - 5 r_0 - r_0^2/4) + (1+ r_0)^3 = 0
\end{equation}
and so the locus of the saddle-node bifurcation only depends on $r_0$
and $z$.  It can also be shown that at $r_0\!=\!0$ always one of the
non trivial fixed points crosses through the fixed point of the origin
in a transcritical bifurcation. The various types of flows that occur
as function of $z$ and $r_0$ are illustrated in Fig.~\ref{figflow}.

\begin{figure*}
\includegraphics[width=16cm]{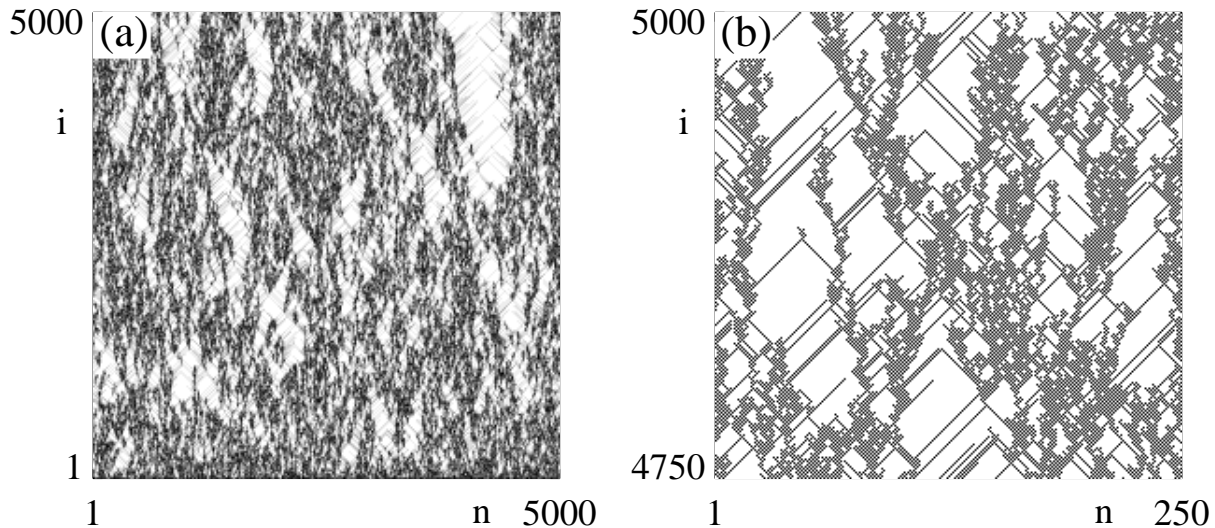}
\caption[]{(a) Large scale dynamics of our stochastic model for $d
=0.01$, $c=0.1$ and $p=0.614$. The grey scale corresponds to the
number of solitons and active sites coarse grained in a cell of 20
space and 20 time units. (b) Blow up of the dynamics shown in the top
left corner of (a).}\label{fig12b}
\end{figure*}

As shown in Fig.~\ref{figflow}, there are essentially four
qualitatively different types of flow and two bifurcations occurring.
We will here discuss these flowtypes and their relevance for the
dynamics. {\em{(a)}} Only the trivial fixed point is present, and is
stable.  Hence all initial conditions flow to the absorbing state.
{\em{ (b)}} For small soliton lifetime $z>1$, the two non-trivial
fixed points $(S_1^*,C_1^*)$ and $(S_2^*,C_2^*)$ that are born in a
saddle-node bifurcation do not lie in the first quadrant and are
therefore not relevant for the mean field equations. Hence the
situation in {\em{(b)}} means that there is a single relevant fixed
point at the origin and so the system is in the absorbing state.
{\em{(c)}} When, for $z>1$, $r_0$ crosses through zero from below,
$(S_1^*,C_1^*)$ crosses through the origin in a transcritical
bifurcation. All initial conditions in the first quadrant flow now to
$(S_1^*,C_1^*)$; the mean field equations indicate that there is a
finite activity, whose value grows approximately linearly in $r_0$. The
transition at $r_0\!=\!0$ corresponds to the standard DP transition for
$z>1$. {\em{(d)}} For long soliton lifetimes ($z<1$) the two
non-trivial fixed points $(S_1^*,C_1^*)$ (square) and $(S_2^*,C_2^*)$
(triangle) are also created in a saddle-node bifurcation; but in
contrast to case (b) both lie in the first quadrant and are therefore
{\em relevant} for the dynamics.  Depending on initial conditions, the
final state can either be absorbing or active; the incoming manifold
of the saddle point acts as a separatrix. The transition that occurs
here as the saddle-node bifurcation is crossed leads to a finite jump
in the value of $c$ in the active state, which is indicative of a
first order transition.  {\em{(e)}} When, for $z<1$, $r_0$ crosses
through zero from below, $(S_2^*,C_2^*)$ crosses through the origin in
a transcritical bifurcation. All initial conditions in the first
quadrant flow now to $(S_2^*,C_2^*)$.

To study the phase transition we shall primarily vary $r_0$ while
keeping $a$ and $u$ fixed. There are three generic choices for $z$
relevant here:

\begin{itemize}
\item $z \rightarrow \infty$. In this case the solitons have
probability 1 to die once they are generated, and so the system is
effectively soliton-free. This is the case of pure DP, and the
transition takes place at $r_0=0$. There is no hysteresis.
\item $z > 1$. This is the regime of short soliton lifetimes.  Here
the solitons do not contribute to any change in the qualitative
behavior. An attractive fixed point $S=S^*_1 \approx \frac{ar}{z-1}$
emerges for small, positive $r_0$. This corresponds to $C=C^*_1
\approx \frac{a^2 r_0}{z-1}$, such that this fixed point converges
towards the DP value $C^*_1 \rightarrow r_0/u$ for large $a$. As $r_0
\rightarrow 0$ this fixed point converges towards the origin and it
changes stability at $r_0\!=\!0$ (see figure \ref{figflow}), implying
that the transition is continuous.  Thus, the transition for small
soliton lifetimes $(z>1)$ still takes place at $r_0\!=\!0$ and
resembles DP.
\item $z < 1$. This the soliton dominated regime where a completely
different scenario occurs.  For $r_0>0$ the behavior is determined by
the stable node at $S^*_2 \approx a(z^{-1/2} -1)$. When $r_0$ becomes
negative this fixed point remains stable and away from the origin.
Simultaneously the origin becomes attractive and a saddle appears
close to the origin at $S=S^*_1 \approx \frac{ar_0}{z-1}$.  Initial
conditions close to the origin will evolve into that point, while
initial conditions above the stable manifold for the saddle located at
$(S^*_1,C^*_1)$ will converge toward the node $(S^*_2,C^*_2)$.  This
will go on until the saddle $(S^*_1,C^*_1)$ and the node
$(S^*_2,C^*_2)$ merge in a saddle-node bifurcation at $r_0=r_c(z)$.
Below this critical point the origin is globally attractive and every
trajectory in the phase space converges towards this.  Going back and
forth along scenarios (a),(d) and (e) there is hysteresis and so for
$z<1$ we clearly observe a first order transition.
\end{itemize}

For infinite soliton lifetimes ($a=0, z=0$) the critical point is
shifted down to $r_0=-1$. Setting $a=0$ into equation (\ref{reduced})
yields the fixed point $S^*=\sqrt{(1+r_0)/u}$ which shows that the
transition is continuous, but with $\beta=1/2$ instead of the DP
mean-field value, $\beta_{DP}=1$.

Finally, at the tri-critical point ($z=1$) equation (\ref{reduced}) is
reduced to $a^2f(S)=S^3+2aS^2-a^2r_0S-a^3r_0$. At $r_0\!=\!0$ the only
non-negative root is $S=0$, but for small positive $r_0$ a new root
appears at $S^*\approx a\sqrt{r_0/2}$. The transition thus remains at
$r_0\!=\!0$ and is continuous, but again with $\beta=1/2$ instead of
the DP value $\beta_{DP}=1$.

\subsection{Phenomenology and statistical properties
of the stochastic model}\label{ssphe}

\begin{figure}[b]
\includegraphics[width=8.5cm]{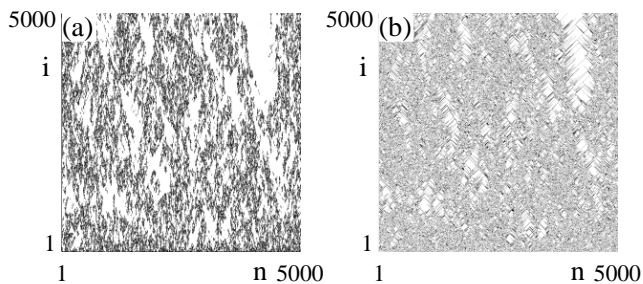}
\caption[]{Concentration of active sites (a) and solitons (b), for
the state shown in Fig.~\ref{fig12b}.}\label{fig12c}
\end{figure}

Let us now discuss the properties of the full stochastic model based
on direct numerical simulations. For small but finite values of the
soliton lifetime ($d \gg 0$) or for sufficiently small production of
solitons ($ c\ll 1$) the transition from inactive to active states
that occurs when $p$ is increased is of second order and indeed
appears to be in the DP universality class. There is, however, also a
regime in which the model, at first glance, appears to display a first
order transition.  In the remainder of the discussion on the
stochastic model we will focus on this regime, which shows some
interesting new features.

The phenomenology of this regime will be illustrated following
Figs.~\ref{fig12b}-\ref{fig12c} where different aspects of the
dynamics of our model are shown. The parameters chosen are
somewhere in the transitory regime, which in the mean field
description corresponds to the regime with two stable fixed points
(Fig.~\ref{figflow}b).

In Fig.~\ref{fig12b}a we show the evolution of our model, starting
from a fully active state.  Fig.~\ref{fig12b}b is a close-up of
the top left corner of Fig.~\ref{fig12b}a which shows the dynamics
of active sites and solitons in detail. At first glance the
clusters of activity look extremely similar to the ordinary
Domany-Kinzel Cellular Automaton, but after closer inspection it
becomes clear that colliding solitons generate new active clusters
(one example can be seen in Fig.~\ref{fig12b}b for $n\approx 50, i
\approx 4825$).

While Fig.~\ref{fig12b} shows both solitons and active sites, we
have shown the coarse grained activity and solitons separately in
Figs.~\ref{fig12c}a-b. Clearly, the soliton density is more
uniformly spread, and one can think of the coarse grained
dynamics as active clusters surrounded by clouds of solitons.

\begin{figure}[t]
\includegraphics[width=8.5cm]{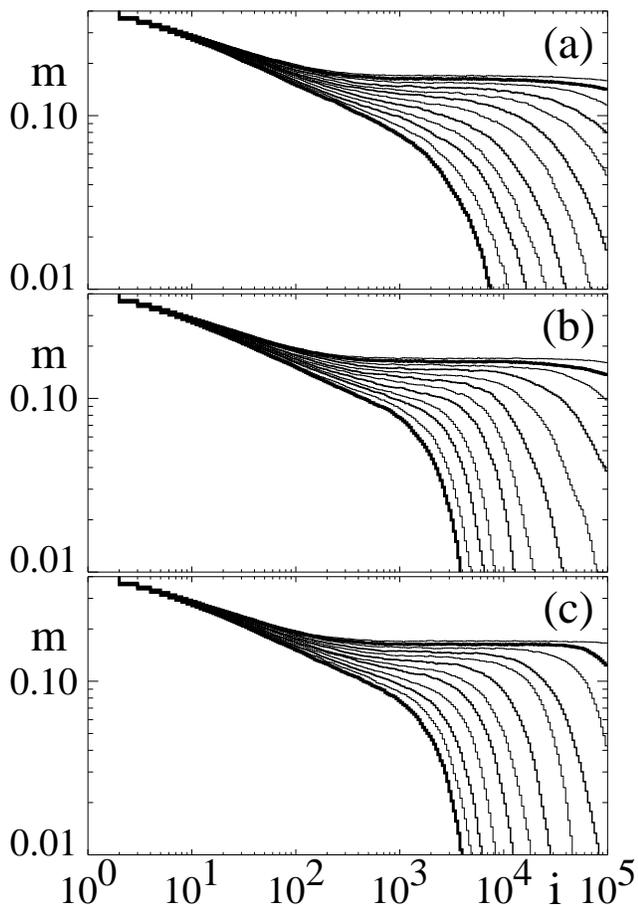}
\caption[]{Decay of average activity $m$ for $c=0.1$, $d=0.01$, and
$p=0.610,0.611,\dots,0.621$ (increasing $p$ leads to an increase of
activity; the curves with $p=0.610$ and $0.620$ are thicker). Averages
are taken over (a) 2000 systems of L=200, (b) 200 systems of L=2000
(c) 20 systems of L=20000.  }\label{figac1}
\end{figure}

To gain insight in the statistical properties of our model, we have
studied the decay of the number of active sites as a function of time,
for a range of system sizes $L$ and parameter values $p$. Here and in
the remainder, we keep the soliton-parameters $d$ and $c$ at values
$0.01$ and $0.1$ respectively.  In Fig.~\ref{figac1} we show the
results of these calculations for $p$ ranging from 0.612 to 0.621.

Clearly, the decay of activity looks quite different from DP. If
we focus on the activity as a function of $p$ for a fixed large
time $\tau$, we find a very abrupt transition at $p = p_1$ from an
inactive to an active state, with a value of the activity given by
the ``plateau'' that can be seen in Fig.~\ref{figac1}. This
behavior is indicative of a first order transition, consistent
with the mean field theory for large soliton lifetimes.

Let us focus on the decay of activity in more detail. When $p$ is
small enough, the activity will decay faster than a power-law. When
$p$ large enough so that active states will spread, we are
certainly above the transition and the activity will reach a
plateau value.  For DP, there is a critical value where the
activity decays as a power-law, but this is not necessarily the case
in our model. As shown in Fig.~\ref{figac1} there is a transient
decay towards the plateau value which can look like a power-law, as
shown in more detail in Fig.~\ref{figac}. For a transient period
that in this case goes up to $t\!\approx\!10^3$, it is possible
to find values of $p$ such that the decay of $m$ appears to be a
power-law with a non-DP exponent. For the example shown, a
reasonable scaling can be obtained over 2 decades. We speculate
that this may be the origin of the non-universal power-laws
observed in coupled map lattices \cite{CM,CM2}. However, for this
to be real asymptotic scaling, one should be able to have this
scaling extend to arbitrary large times; however the activity
curves for sizes 200, 2000 and 20000 all bend downwards at nearly
the same time; hence there is no hope that increasing the
systemsize extends the time interval over which apparent scaling
can be found. For long times the activity either decays rapidly,
or first hits a plateau. Clearly, the transition is not an
ordinary 2nd order transition.

\begin{figure}
\includegraphics[width=8.5cm]{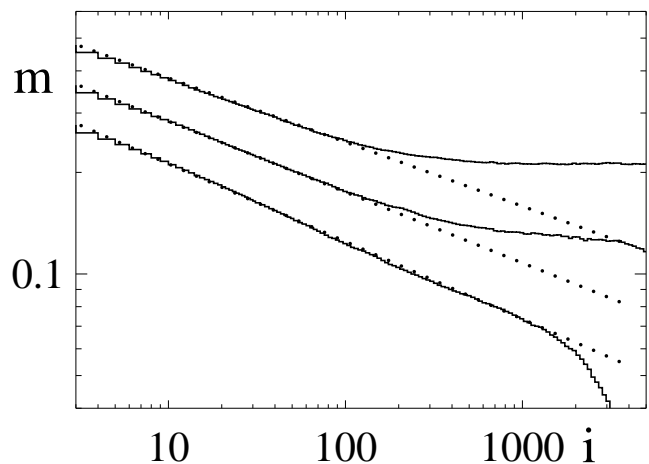}
\caption[]{Average activity $m$ for $c=0.1$, $d=0.01$ in an ensemble of
$20$ systems of $L=20000$, showing the appearance of quasi-power-law
decay. The three curves correspond to $p=0.612, 0.616$, and $p= 0.62$
respectively, and are shifted by 30\% for clarity. The three straight
lines corresponds to power-laws with exponents $-0.19, -0.21$ and
$-0.23$. In particular the scaling in the system for $p=0.612$ looks
rather convincing with an exponent of $-0.23$.  }\label{figjoke}
\end{figure}

Is this transition now an ordinary 1st order transition? When we
increase the time scale $\tau$ where we inspect the average
activity, the transition becomes sharper, and the transition value
$p_1$ will shift. We do not see evidence for $p_1$ going to a
well-defined asymptotic value when $\tau$ goes to infinity, at
least not on the timescales that we can probe numerically, and in
this sense the transition is not truly first order.

We will argue now that the plateau indeed does not represent the
truly asymptotic behavior.  Let us return to Fig.~\ref{fig12c},
where the activity appears to arrive at the plateau (the overall
activity appears to approach a constant). However, around
$n\approx 4000, i \approx 3000$ a large ``hole'' opens up. Once
the size of this hole becomes larger than twice the lifetime of
the solitons, it becomes unlikely that colliding solitons will
create new activity there and ``heal '' the hole. In fact, for
this particular example the hole did spread out and the system
decayed to the inactive state. A closer inspection of the
dynamical states that occur when the activity drops below the
plateau value, shows that this is the general scenario: the
nucleation and subsequent spreading of a large inactive droplet is
what dominates the asymptotic decay of the active states here.

\begin{figure}[b]
\includegraphics[width=8.5cm]{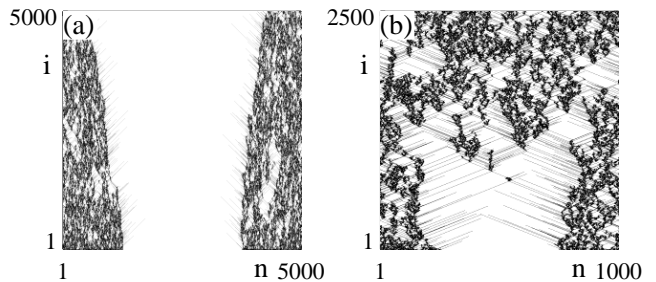}
\caption[]{Dynamics for $c=0.1,d=0.01$ and $p=0.618$, showing that a
hole of size 500 is healed (a) while a larger hole of size 2500 is
{\em not} healed (b). }\label{fignotheal}
\end{figure}

The spreading or shrinking of a large hole is governed by the
propagation velocity of the coarse grained domain-walls between
active and inactive patches. What we have found is that the
apparent first order transition point $p_1$ does not coincide with
the point $p^*$ where this velocity changes sign.  We illustrate
this property by following the dynamics of a large inactive
droplet for $p=0.618$, where the system has a well defined plateau
in the activity (see Fig.~\ref{figac1}). Thus $p>p_1$, but
nevertheless, a hole of size 5000 grows as can be seen in
Fig.~\ref{fignotheal}.  A hole of size 500 is healed for these
same parameter values, as shown in Fig.~\ref{fignotheal}.

The solitons in our model thus introduce an additional length-scale
of order $1/d$, and as a result the size of the {\em inactive}
holes becomes of importance. When such holes are smaller than
$2/d$, solitons will penetrate deeply enough to collide and create
new patches of activity within this hole. Since the first order
transition is triggered by the solitons, this indicates that the
velocity of the domain wall is not strongly dependent on these
degrees of freedom. This is in accord to the results from the mean
field analysis, if we assume that the properties of the domain
walls are still governed by the trivial fixed point with
$C\!=\!0$.

The difference between the spreading of a small active cluster and the
behavior of an homogeneously active state indicates that the initial
concentration of active sites plays a role. This is illustrated in
Fig.~\ref{figac} where we follow the evolution of the activity for a
range of initial concentrations of activity for $p\!=\!0.621$.  For
initial activities in the range from 1 to 0.1, the same plateau value
is reached, but for initial activities of $0.05$ and smaller, there is
an initial increase of the activity after which the activity rapidly
decays; the plateau is never reached.

\begin{figure}
\includegraphics[width=8.5cm]{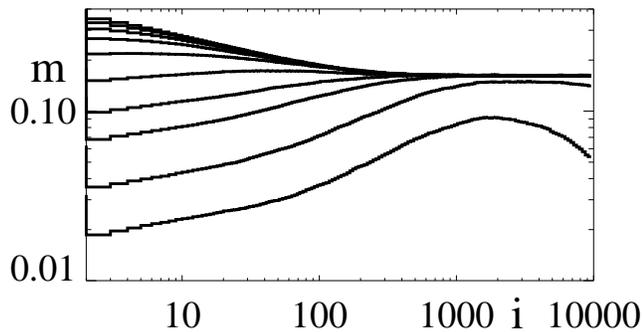}
\caption[]{Evolution of the average activity $m$ in 20 systems of size
20000 for $c=0.1, d=0.01$ and $p=0.621$, i.e., well in plateau regime.
Here we vary the initial concentration of active states by randomly
distributing active sites through our lattice for $i=1$. These
activities are, respectively, $1.0,
0.85,0.7,0.55,0.4,0.25,0.15,0.1,0.05$ and $0.025$. For the latter two
cases, the plateau is not reached, even though initially the activity
is increasing. That the long time behavior ($i>200$) depends on the
initial concentration is reminiscent of a first order
transition.}\label{figac}
\end{figure}

\begin{figure}[t]
\includegraphics[width=8.5cm]{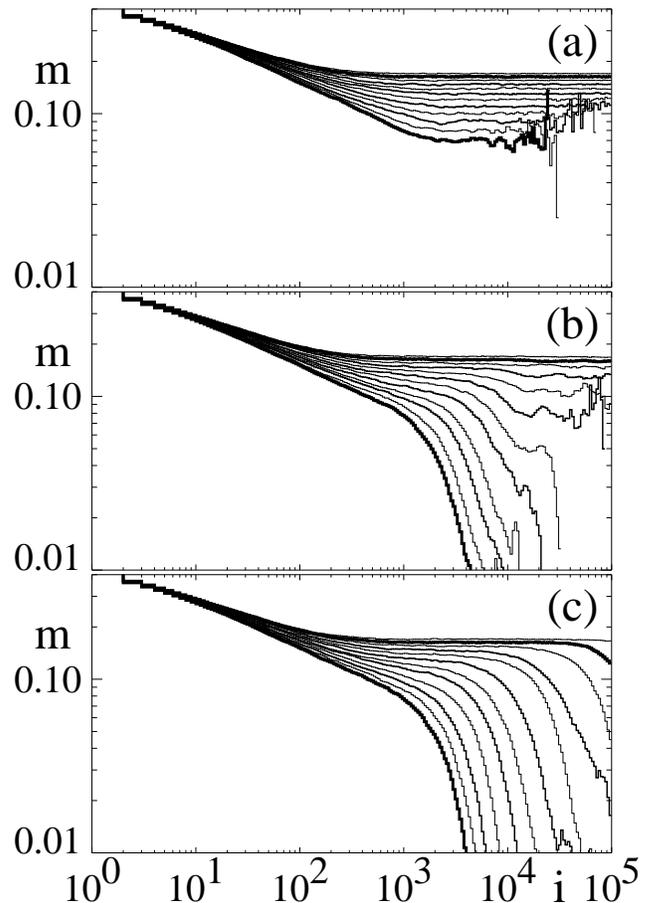}
\caption[]{Average activity divided by the number of active systems,
for the same parameter values as shown in Fig.~\ref{figac1}.
}\label{figacn}
\end{figure}

In Fig.~\ref{figacn} we show the evolution of the activity $m$ divided
by the number of surviving clusters for the same parameter values.
For small systems these plots are very different from the ones
averaged over all systems (Fig.~\ref{figac1}); in the present case
there is a typical activity in each system which rapidly disappears.
We interpret this as further evidence that the nucleation of large
holes dominates the eventual decay. For larger systems this effect
disappears because the time it takes for a hole to engulf the whole
system is large.

\section{Discussion}\label{disc}

The overall picture that emerges from our study is that the transition
to spatio-temporal intermittency is strongly influenced by coherent
ballistically traveling ``solitons", which, even though they have a
finite life-time, change the nature of the transition and can
introduce first order like behavior. That such a scenario is relevant
is supported by recent evidence for a discontinuous transition to
spatio-temporal chaos in the damped Kuramoto-Sivashinsky equation
\cite{elder}, which is well-known to support localized ballistically
moving excitations, or ``pulses" \cite{balmforth}.

We build our conclusions upon an extension, using two dimensional
local maps, of the Chat\'e-Manneville coupled map lattice. We thereby
gain an additional parameter, which turns out to tune the importance
and life-time of the solitons. For this coupled map lattice we find,
depending on parameters, evidence for both continuous phase
transitions in the universality class of Directed Percolation with
infinitely many absorbing states and for first order behavior.

To understand this behavior, we have developed a stochastic model
generalizing the Domany-Kinzel cellular automaton. In this model, the
active sites can emit solitons and by colliding, the solitons can
create new active sites. Simulations of this model, together with the
appropriate mean field theory, support the existence of both
continuous and discontinuous transitions. With the stochastic model
one can look at the behavior on much larger length and time
scales. One thereby discovers that the active states close above the
discontinuous transitions are actually metastable, and will finally
decay when a sufficiently large droplet nucleates. Such breakdown of
first order behavior has been predicted by Hinrichsen
\cite{hinrich}. We have not been able to pin down exactly how the
nucleation time and eventual decay depends on system size, but it is
not ruled out that the truly asymptotic behavior is still given by
Directed Percolation in accordance with the predictions of Hinrichsen.

Even though the first order behavior is probably not truly
asymptotic, it appears very clearly over a surprisingly long range of
intermediate time scales, and would thus be relevant in the
interpretation of experiments. We further show that this feature can
lead to long power like transients displaying non-universal ``critical
exponents," and we believe that such transients are the origin of the
observed non-universality in the transition to spatio-temporal
intermittency.

\end{document}